\pdfoutput=1
\RequirePackage{ifpdf}
\ifpdf 
\documentclass[pdftex]{sigma}
\else
\documentclass{sigma}
\fi

\begin{document}

\allowdisplaybreaks

\renewcommand{\PaperNumber}{012}

\FirstPageHeading

\ShortArticleName{Geometric Constructions Underlying Relativistic Description of Spin}

\ArticleName{Geometric Constructions Underlying Relativistic\\
Description of Spin on the Base of Non-Grassmann\\
Vector-Like Variable}

\Author{Alexei A.~DERIGLAZOV and Andrey M.~PUPASOV-MAKSIMOV}

\AuthorNameForHeading{A.A.~Deriglazov and A.M.~Pupasov-Maksimov}

\Address{Departamento de Matem\'atica, ICE, Universidade Federal de Juiz de Fora, MG, Brasil}
\Email{\href{mailto:alexei.deriglazov@ufjf.edu.br}{alexei.deriglazov@ufjf.edu.br},
\href{mailto:pupasov@phys.tsu.ru}{pupasov@phys.tsu.ru}}

\ArticleDates{Received December 17, 2013, in f\/inal form February 04, 2014; Published online February 08, 2014}

\Abstract{Basic notions of Dirac theory of constrained systems have their analogs in dif\/ferential
geometry. Combination of the two approaches gives more clear understanding of both classical and quantum mechanics,
when we deal with a~model with complicated structure of constraints.
In this work we describe and discuss the spin f\/iber bundle which appeared in various mechanical models
where spin is described by vector-like variable.}


\Classification{53B50; 81R05; 81S05}

\section{Introduction}

Already in 1926, Frenkel observed~\cite{Frenkel, Frenkel2} that variational formulation of relativistic
spinning particle represents a~nontrivial problem, if we intend to take into account the conditions which
guarantee the right number of degrees of freedom in the theory.
This leads to a~rather complex singular Lagrangians, see~\cite{bbra,bt,berezin:1977,cognola1981lagrangian,corben:1968, AAD6,AAD5, Alexei,
GG1, grassberger1978, hanson1974relativistic} and references
therein.
In the Hamiltonian formulation, this implies a~theory with Dirac's f\/irst and second-class constraints.
Basic notions of the Dirac theory have their analogs in dif\/ferential geometry.
Second-class constraints mean that all true trajectories lie on a~curved submanifold of the initial
phase-space.
The Dirac bracket, constructed on the base of second-class constraints, induces canonical symplectic
structure on the submanifold.
Besides, due to the f\/irst-class constraints (equivalently, due to local symmetries), a~part of variables
have ambiguous evolution.
This also can be translated into geometric language: due to the ambiguity, the submanifold is endowed with
a~natural structure of f\/iber bundle.
Physical variables are functions of the coordinates which parameterize the base of the f\/iber bundle.

Here we discuss the models where spin is considered as a~composed quantity (inner angular momentum)
constructed from non-Grassmann vector-like variable and its conjugated momentum~\cite{cognola1981lagrangian,deriglazov-ns:2010, AAD3, AAD6, AAD5, Alexei, AAD4,grassberger1978,peletminski2005lagrangianspin}.
They lead to the spin f\/iber bundle shortly described in~\cite{deriglazov-ns:2010, AAD3, AAD4}.
In the present work we start from non-relativistic spinning particle~\cite{deriglazov-ns:2010} and describe
the corresponding non-relativistic bundle in a~more systematic form.
Then we represent this in a~manifestly Poincare-covariant form, arriving at the set of constraints which
guarantees those of Frenkel~\cite{Frenkel} and Bargmann, Michel and Telegdi (BMT)~\cite{BMT} theories.
The passage from vector-like to spin variable is not a~change of variables, and acquires a~natural
interpretation in this geometric construction.
This turns out to be useful in the study of both classical and quantum mechanics of a~spinning
particle~\cite{Alexei, DPW1,laroze2009spin-gravity, DPW2}.

\newpage

\section{Non relativistic spin}

 \subsection{Basic constraints}\label{subsec:intro-NRsp}

 The Pauli equation involves spin operators which are proportional to the Pauli matrices, $\hat
S_i=\frac{\hbar}{2}\sigma_i$.
They form ${\rm SO}(3)$-algebra with respect to commutator
\begin{gather}
\label{intr.5}
[\hat S_i,\hat S_j]_{-}=i\hbar\epsilon_{ijk}\hat S_k.
\end{gather}
Here $\epsilon_{ijk}$ is totally antisymmetric tensor with $\epsilon_{123}=1$.
Besides, they obey the identity
\begin{gather}
\label{intr.7}
\hat{\bf S}{}^2=\hbar^2s(s+1)=\frac{3\hbar^2}{4},
\end{gather}
which corresponds to the value of ${\rm SO}(3)$-Casimir $s=\frac{1}{2}$.

To construct the corresponding semiclassical model, we look for a~classical-mechanics system which, besides
the position variables $x_i$, contains additional degrees of freedom, appropriate for the description of
a~spin: in the Hamiltonian formulation the spin should be described, at the end, by three variables $S_i$
which imply~\eqref{intr.5} and~\eqref{intr.7} in the course of canonical quantization.
According to this, typical spinning-particle model consist of a~point on a~world-line and some set of
variables describing the spin degrees of freedom, which form the inner space attached to that
point~\cite{hanson1974relativistic}.
In fact, dif\/ferent spinning particles discussed in the literature dif\/fer by the choice of inner space
of spin.
Exceptional case is the rigid particle which consist of only position variables, but with the action
containing higher derivatives.
Recently it has been shown~\cite{deriglazov2013rigid}, that this yields the Dirac equation, hence it can be
used for description of spin.

We intend to construct spinning particle starting from a~proper variational problem.
This is the f\/irst task we need to solve, as the formulation of a~variational problem in closed form is
known only for the case of a~phase space equipped with canonical Poisson bracket, say $\{\omega_i,
\pi_j\}=\delta_{ij}$.
The number of variables and their algebra are dif\/ferent from the number of spin operators and their
commutators,~\eqref{intr.5}.
To improve this, we need to impose constraints and/or to pass from the initial to some composed variables.
This implies the use of Dirac machinery for constrained theories~\cite{dirac1950lectures}.

The most natural way to to arrive at the operator algebra~\eqref{intr.5} is to consider spin variables as
composed quantities,
\begin{gather}
\label{intr.9.1}
S_i=\epsilon_{ijk}\omega_j\pi_k,
\end{gather}
where ${\boldsymbol\omega}$, $\boldsymbol\pi$ are coordinates of a~phase space equipped with canonical
Poisson bracket.
This immediately induces ${\rm SO}(3)$-algebra for $S_i$
\begin{gather*}
\{S_i(\omega,\pi),S_j(\omega,\pi)\}_{\rm PB}=\epsilon_{ijk}S_k.
\end{gather*}
Unfortunately, this is not the whole story.
First, we need some mechanism which explains why~${\bf S}$, not $\boldsymbol{\omega}$ and
$\boldsymbol{\pi}$ must be taken for the description of spin degrees of freedom.
Second, the basic space is six-dimensional, while the spin manifold is two-dimensional (we remind that the
square of spin operator has f\/ixed value, equation~\eqref{intr.7}).
To improve this, we look for variational problem which, besides dynamical equations, implies the
constraints ($a$ and $b$ are given numbers)
\begin{gather}
\label{intr.11}
{\boldsymbol{\omega}}^2=a^2,
\qquad
{\boldsymbol{\pi}}^2=b^2,
\qquad
{\boldsymbol{\omega}}{\boldsymbol{\pi}}=0.
\end{gather}
Then
\begin{gather}
\label{intr.12}
{\bf S}^2={\boldsymbol{\omega}}^2{\boldsymbol{\pi}}^2-({\boldsymbol{\omega}}{\boldsymbol{\pi}}
)^2=\frac{3\hbar^2}{4},
\end{gather}
if we put $b^2=\frac{3\hbar^2}{4a^2}$.
We point out that this is essentially unique ${\rm SO}(3)$-invariant three-dimensional surface of ${\mathbb
R}^6$.
The same result~\eqref{intr.12} follows from
\begin{gather}
\label{intr12.1}
\boldsymbol\omega\boldsymbol\pi=0,
\qquad
\boldsymbol\pi^2-\frac{a}{\boldsymbol\omega^2}=0,
\qquad
a=\frac{3\hbar^2}{4}.
\end{gather}
These constraints naturally arise in the model of rigid particle, see~\cite{deriglazov2013rigid}.
Below we discuss Lorentz-covariant description for both sets.

While ${\bf S}$ in~\eqref{intr.9.1} looks like angular momentum, the crucial dif\/ference with orbital
angular momentum is the prsence of local symmetry, which acts on the basic variables $\vec{\omega}$,
$\vec{\pi}$, while leaves invariant the spin variable~${\vec{S}}$.
We refer this as spin-plane symmetry.
Using analogy with classical electrodynamics, $\boldsymbol{\omega}$ and $\boldsymbol{\pi}$ are similar to
four-potential $A^\mu$ while ${\bf{S}}$ plays the role of~$F^{\mu\nu}$.
According to the general theory~\cite{deriglazov2010classical, dirac1950lectures,gitman1990quantization}, in
this case the coordinates $\boldsymbol{\omega}$ of the ``inner-space particle'' are not physical
(observable) quantities.
The only observable quantities are the gauge-invariant variables~${\bf{S}}$.
So our construction realizes, in a~systematic form, the oldest idea about spin as ``inner angular momentum''.

\subsection{Spin-sector Lagrangian and Hamiltonian}

As the Lagrangian which implies the constraints~\eqref{intr.11}, we take the expression $L_{\rm
spin}=\frac{1}{2g}{\dot{\boldsymbol{\omega}}^2}+\frac12gb^2-\frac{1}{\phi}\big(\boldsymbol{\omega}^2-a^2\big)$,
where $\boldsymbol{\omega}=(\omega_1, \omega_2, \omega_3)$.
Variation with respect to auxiliary variables~$g(t)$ and~$\phi(t)$ gives the equations~$\dot{\boldsymbol{\omega}}^2=g^2b^2$ and~$\boldsymbol{\omega}^2=a^2$, the latter implies~$\dot{\boldsymbol{\omega}}\boldsymbol{\omega}=0$.
In the Hamiltonian formulation these equations turn into the desired constraints.
We can integrate out the variable~$g$, presenting Lagrangian in more compact form
\begin{gather*}
L_{\rm spin}=b\sqrt{{\dot{\boldsymbol{\omega}}}^2}-\frac{1}{\phi}\big(\boldsymbol{\omega}^2-a^2\big).
\end{gather*}
This also gives the desired constraints.
The last term represents kinematic (velocity-inde\-pen\-dent) constraint which is well known from classical
mechanics.
So, we might follow the classical-mechanics prescription to exclude $\phi$ as well.
But this would lead to lose of manifest rotational invariance of the formalism.
The model is manifestly invariant under rotations $R_{ij}$.
The auxilia\-ry variable $\phi$ is a~scalar under these transformations.
For the $\omega_i$ we have $\omega'_i=R_{ij}\omega_j$.
There is also local (spin-plane) symmetry with the parameter $\beta(t)$
\begin{gather}
\label{nnr3.2}
\delta\omega_i=\beta\dot\omega_i,
\qquad
\delta\phi=\frac{1}{\phi^2}\left(\frac{\beta}{\phi}\right)^{\textstyle\cdot}.
\end{gather}
Let us consider Hamiltonian formulation of the model.
Equations for the conjugate momenta~$\pi_i$ read~$\pi_i=b\frac{\dot\omega_i}{\sqrt{\dot{\boldsymbol
\omega}^2}}$.
This implies the primary constraint $\boldsymbol{\pi}^2=b^2$.
Momentum for $\phi$ turns out to be one more primary constraint, $\pi_\phi=0$.
The complete Hamiltonian reads
\begin{gather}
\label{nnr6}
H=\frac{\lambda_1}{2}\big(\boldsymbol{\pi}^2-b^2\big)+\frac{1}{\phi}\big(\boldsymbol{\omega}^2-a^2\big)+\lambda_2\pi_\phi.
\end{gather}
We have denoted by $\lambda_a$ the Lagrangian multipliers for the primary constraints.
Equation~\eqref{nnr3.2} induces~\cite{deriglazov2000local} the inf\/initesimal phase-space transformations
\begin{gather*}
\delta\omega_i=\phi\beta\pi_i,
\qquad
\delta\pi_i=-\frac{2}{\lambda_1}\beta\omega_i,
\\
\delta\phi=\phi^2\left(\frac{\beta}{\lambda_1}\right)^{\textstyle\cdot},
\qquad
\delta\pi_\phi=0,
\qquad
\delta\lambda_1=-(\phi\beta)^{\textstyle\cdot},
\qquad
\delta\lambda_2=-(\delta\phi)^{\textstyle\cdot}.
\end{gather*}
They leave invariant the Hamiltonian action
\begin{gather}
\label{nnr3.3}
S_H=\int {\rm d}t\pi\dot\omega+\pi_\phi\dot\phi-H.
\end{gather}
Finite transformations and their geometric interpretation will be discussed in
Section~\ref{subsec:so3FB}.

Applying the Dirac procedure, we obtain the following sequence of constraints and equations for the
Lagrangian multipliers: $\pi_\phi=0$  $\Rightarrow$ $\boldsymbol{\omega}^2-a^2=0$
$\Rightarrow$ $(\boldsymbol{\omega}\boldsymbol{\pi})=0$  $\Rightarrow$ $\lambda_1=-\frac{2a^2}{b^2\phi}$.
Hence all the desired constraints~\eqref{intr.11} appeared.
Lagrangian multiplier $\lambda_1$ can be substituted into~\eqref{nnr6}.
Besides the constraints, the Hamiltonian~\eqref{nnr6} implies the dynamical equations
$\dot\phi=\lambda_\phi$, $\dot\pi_\phi=0$, $\dot\omega_i=-\frac{2a^2}{b^2\phi}\pi_i$, and
$\dot\pi_i=\frac{2}{\phi}\omega_i$.
Neither equations nor constraints determine the variables $\lambda_\phi$ and $\phi$, the latter enters as
an arbitrary function into general solution for the variables $\boldsymbol{\omega}$ and $\boldsymbol{\pi}$.
Hence, the dynamics of these variables is ambiguous.
This is in correspondence with invariance of the action~\eqref{nr3} under local
transformations~\eqref{nnr3.2}.
According to general theory~\cite{deriglazov2010classical, dirac1950lectures,gitman1990quantization},
$\boldsymbol{\omega}$ and $\boldsymbol{\pi}$ are not observable quantities.
The variables $S_i=\epsilon_{ijk}\omega_j\pi_k$ have unambiguous evolution, as it should be $\dot S_i=0$.
In interacting theory ${\bf S}$ will precess under torque exercised by magnetic f\/ield, see below.
Due to equations~\eqref{intr.11}, the coordinates $S_i$ obey~\eqref{intr.12}.

\subsection[${\rm SO}(3)$ spin surface and associated spin f\/iber bundle]{$\boldsymbol{{\rm SO}(3)}$ spin surface
and associated spin f\/iber bundle}\label{subsec:so3FB}

While our model~\eqref{nr3} consists of the basic variables $\boldsymbol{\omega}$ and $\boldsymbol{\pi}$,
quantum mechanics obtained in terms of spin variables ${\bf{S}}$.
The passage from $\boldsymbol{\omega}$, $\boldsymbol{\pi}$ to ${\bf{S}}$ is not a~change of variables, and
acquires a~natural interpretation in the geometric construction described below.
This can be resumed as follows.
All the trajectories $\boldsymbol{\omega}(t)$, $\boldsymbol{\pi}(t)$ belong to the surface~\eqref{intr.11}
of phase space which we identify with the group manifold ${\rm SO}(3)$.
The map $(\boldsymbol{\omega}, \boldsymbol{\pi}) \rightarrow {\bf{S}}$ determines natural structure of
f\/iber bundle on ${\rm SO}(3)$ with the structure group being ${\rm SO}(2)$.
The structure group turn into local symmetry in dynamical theory and selects ${\bf{S}}$ as the physical
(observable) variables.
Canonical quantization of the f\/iber bundle yields the Pauli equation, see the next section.
Generalization on ${\rm SO}(k, n)$ Lie--Poisson manifold has been presented
in~\cite{deriglazov2012variational}.

In this section we normalize the basic spin variables in such a~way, that the constraints have the form
$\boldsymbol{\omega}^2-1=0$, $\boldsymbol{\pi}^2-1=0$ and $(\boldsymbol{\omega}\boldsymbol{\pi})=0$.

Consider six-dimensional phase space equipped with canonical Poisson bracket
\begin{gather*}
\mathbb{R}^{6}=\{\omega_i,\pi_j;\{\omega_i,\pi_j\}_{\rm PB}=\delta_{ij}\},
\end{gather*}
and three-dimensional spin space $\mathbb{R}^3=\{S_i\}$.
Def\/ine the map
\begin{gather}
f: \ \mathbb{R}^{6}~\rightarrow~\mathbb{R}^{3},
\qquad
f: \ (\omega_i,\pi_j)~\rightarrow~S_i=\epsilon_{ijk}\omega_j\pi_k,
\nonumber
\\
\text{or}
\qquad
{\bf S}=\boldsymbol{\omega}\times\boldsymbol{\pi},
\qquad
\text{rank}\frac{\partial(S_i)}{\partial(\omega_j,\pi_k)}=3.
\label{ss.2}
\end{gather}
Poisson bracket on $\mathbb{R}^{6}$ and the map~\eqref{ss.2} induce ${\rm SO}(3)$ Lie--Poisson bracket on
$\mathbb{R}^3$
\begin{gather*}
\{S_i,S_j\}\equiv\{S_i(\omega,\pi),S_j(\omega,\pi)\}_{\rm PB},
\qquad
\{S_i,S_j\}=\epsilon_{ijk}S_k.
\end{gather*}
According to previous section, all trajectories $\boldsymbol{\omega}(t), \boldsymbol{\pi}(t)$ lie on ${\rm
SO}(3)$-invariant surface of $\mathbb{R}^{6}$
\begin{gather}
\label{ss.4}
\mathbb{T}^3=\{\boldsymbol{\omega}^2-1=0,
\,
\boldsymbol{\pi}^2-1=0,
\,
\boldsymbol{\omega}\boldsymbol{\pi}=0\}.
\end{gather}
 $\mathbb{T}^3$ can be identif\/ied with group manifold ${\rm SO}(3)$.
Indeed, given $\boldsymbol{\omega}$, $\boldsymbol{\pi}$, consider $3\times 3$ matrix with the lines
$\boldsymbol{\omega}$, $\boldsymbol{\pi}$ and $\boldsymbol{\omega}\times\boldsymbol{\pi}$
\begin{gather}
\label{ss.5}
R=\left(
\begin{matrix}
\boldsymbol{\omega}
\\
\boldsymbol{\pi}
\\
\boldsymbol{\omega}\times\boldsymbol{\pi}
\end{matrix}
\right).
\end{gather}
Equations~\eqref{ss.4} imply $RR^{\rm T}=1$ and $\det R=1$.
The map $\mathbb{T}^3 \rightarrow {\rm SO}(3)$ given by equation~\eqref{ss.5} determines dif\/feomorphism
of the manifolds.

When $(\boldsymbol{\omega}, \boldsymbol{\pi})\in\mathbb{T}^3$, we have ${\bf
S}^2=\boldsymbol{\omega}^2\boldsymbol{\pi}^2-(\boldsymbol{\omega}\boldsymbol{\pi})^2=1$.
So, $f$ maps the manifold $\mathbb{T}^3$ onto two-dimensional sphere of unit radius (spin surface),
$f(\mathbb{T}^3)=\mathbb{S}^2$.
Denote $\mathbb{F}_S\in\mathbb{T}^3$ preimage of a~point ${\bf{S}}\in\mathbb{S}^2$,
$\mathbb{F}_S=f^{-1}({\bf{S}})$.
This set is composed by all pairs $(\boldsymbol{\omega}, \boldsymbol{\pi})$ which lie on the same plane and
thus related by ${\rm SO}(2)$ rotations of the plane.

The manifold $\mathbb{T}^3$ acquires natural structure of f\/iber bundle
\begin{gather}
\label{ss.7}
\mathbb{T}^3=\big(\mathbb{S}^2,\mathbb{F},f\big),
\end{gather}
with base $\mathbb{S}^2$, standard f\/iber $\mathbb{F}$, projection map $f$ and structure group ${\rm
SO}(2)$.

Transformations of structure group read
\begin{gather}
\label{ss.8}
\begin{array}{@{}l}
\boldsymbol{\omega}'=\boldsymbol{\omega}\cos\beta+\boldsymbol{\pi}\sin\beta,
\\
\boldsymbol{\pi}'=-\boldsymbol{\omega}\sin\beta+\boldsymbol{\pi}\cos\beta
\end{array}
\qquad
\Rightarrow
\qquad
\begin{cases}
\delta\omega_i=\beta\pi_i,
\\
\delta\pi_i=-\beta\omega_i.
\end{cases}
\end{gather}
By construction, they leave inert points of base, $\delta S_i=0$.

In the dynamical realization of previous section, the structure group acts independently at each instance
of time and turn into the local spin-plane symmetry.
To see this, let us present the action functional equivalent to~\eqref{nnr3.3} and invariant
under~\eqref{ss.8}.
Consider extended Hamiltonian action $S_{\rm ext}$ for~\eqref{nnr3.3}.
This obtained adding all the higher-stage constraints (with their own Lagrangian multipliers) to $S_H$.
For the present case this is
\begin{gather*}
S_{\rm ext}=\int {\rm d}t P\dot Q-(H+\lambda_3\boldsymbol{\omega}\boldsymbol{\pi}).
\end{gather*}
It is known~\cite{deriglazov2010classical, gitman1990quantization} that the theories $S_H$ and $S_{\rm ext}$ are
equivalent.
 $S_{\rm ext}$ is invariant (modulo to total derivative) under local version of the
transformation~\eqref{ss.8}, $\beta\rightarrow\beta(t)$, accompanied by the following transformation of
auxiliary variables
\begin{gather}
\label{ss.10}
g'_{ab}=\big(KgK^{\rm T}\big)_{ab}+\frac12\dot\beta\delta_{ab},
\qquad
\lambda'_2=-(\phi')^{\textstyle\cdot},
\end{gather}
where
\begin{gather}
\label{ss.11}
g=\left(
\begin{matrix}
\dfrac{1}{\phi}&\dfrac{\lambda_3}{2}
\vspace{1mm}\\
\dfrac{\lambda_3}{2}&\dfrac{\lambda_1}{2}
\end{matrix}
\right),
\qquad
K=\left(
\begin{matrix}
\cos\beta & \sin\beta
\\
-\sin\beta & \cos\beta
\end{matrix}
\right).
\end{gather}

Let $(\boldsymbol{\omega}, \boldsymbol{\pi})\in\mathbb{T}^3$, $\omega_3\ne 0$.
As local coordinates of $\mathbb{T}^3$ in vicinity of this point, we can take $S_1(\boldsymbol{\omega},\boldsymbol{\pi})$,
$S_2(\boldsymbol{\omega}, \boldsymbol{\pi})$, and~$\omega_3$.
The coordinates adjusted with structure of f\/ibration.
That is~$S_1$,~$S_2$ parameterize the base $\mathbb{S}^2$ while $\omega_3$ parameterizes the f\/iber~$\mathbb{F}$.
The spin-plane symmetry determines physical sector of the theory, and hence play the fundamental role in
this construction, see discussion at the end of Section~\ref{subsec:intro-NRsp}.

\subsection{Canonical quantization and Pauli equation}

To test our procedure, we discuss spinning particle corresponding to the Pauli equation on a~stationary
magnetic f\/ield\footnote{The case of an arbitrary electromagnetic background~\cite{foldy:1978} represents
much more complex problem~\cite{DPW1}.}.
Consider the action
\begin{gather}
\label{nr3}
S=\int {\rm d}t\left[\frac{m}{2}\dot{\bf{x}}^2+\frac{e}{c}{\bf{A}}\dot{\bf{x}}+b\sqrt{(D\boldsymbol{\omega})^2}
-\frac{1}{\phi}\big(\boldsymbol{\omega}^2-a^2\big)\right],
\qquad
D\omega_i=\dot\omega_i-\frac{\mu e}{mc}\epsilon_{ijk}\omega_jB_k.
\end{gather}
The conf\/iguration-space variables are $x_i(t)$, $\omega_i(t)$ and $\phi(t)$.
Here $x_i$ represents the spatial coordinates of the particle with the mass~$m$, the charge~$e$, and
magnetic moment~$\mu$, ${\bf{B}}= {\boldsymbol{\nabla}}\times{\bf{A}}$.
Second term in equation~\eqref{nr3} represent minimal interaction with the vector potential
${\bf{A}}({\bf{x}})$ of an external electromagnetic f\/ield, while the third term contains interaction of
spin with a~magnetic f\/ield.
At the end, it produces the Pauli term in quantum-mechanical Hamiltonian.

Let us construct Hamiltonian formulation for the model.
Equations for the conjugated momenta $p_i$ and $\pi_i$ reads
\begin{gather}
p_i=m\dot x_i+\frac{e}{c}A_i
\qquad
\Rightarrow
\qquad
\dot x_i=\frac{1}{m}\left(p_i-\frac{e}{c}A_i\right),\qquad
\pi_i=b\frac{D\omega_i}{\sqrt{(D\boldsymbol\omega)^2}}.\label{nr5}
\end{gather}
Equation~\eqref{nr5} implies the primary constraint $\boldsymbol{\pi}^2=b^2$.
Momentum for $\phi$ turns out to be one more primary constraint, $\pi_\phi=0$.
The complete Hamiltonian, $H=P\dot Q-L+\lambda_a\Phi_a$, $Q=({\bf{x}}, \boldsymbol{\omega}, \phi)$,
$P=({\bf{p}}, \boldsymbol{\pi}, \pi_\phi)$, reads
\begin{gather}
\label{nr6}
H=\frac{1}{2m}\left(p_i-\frac{e}{c}A_i\right)^2-\frac{\mu e}{mc}\epsilon_{ijk}B_i\omega_j\pi_k+\frac{1}{\phi}
\big(\boldsymbol{\omega}^2-a^2\big)+\frac{\lambda_1}{2}\big(\boldsymbol{\pi}^2-b^2\big)+\lambda_2\pi_\phi.
\end{gather}
We have denoted by $\lambda_a$ the Lagrangian multipliers for the primary constraints
$\Phi_a=(\boldsymbol{\pi}^2-b^2, \pi_\phi)$.

Applying the Dirac procedure, we obtain the following sequence of constraints and equations for the
Lagrangian multipliers
\begin{gather*}
\left.
\begin{matrix}
\boldsymbol{\pi}^2-b^2=0
\\
\pi_\phi=0
\quad
\Rightarrow
\quad
\boldsymbol{\omega}^2-a^2=0
\end{matrix}
\right\}
\qquad
\Rightarrow
\qquad
(\boldsymbol{\omega}\boldsymbol{\pi})=0
\qquad
\Rightarrow
\qquad
\lambda_1=-\frac{2a^2}{b^2\phi}.
\end{gather*}
Lagrangian multiplier $\lambda_1$ can be substituted into~\eqref{nr6}.
Besides the algebraic equations, the Hamiltonian~\eqref{nr6} implies the dynamical equations
\begin{gather}
\dot\phi=\lambda_\phi,
\qquad
\dot\pi_\phi=0,\nonumber
\\
\dot x_i=\frac{1}{m}\left(p_i-\frac{e}{c}A_i\right),
\qquad
\dot p_i=\frac{e}{c}\dot x_j\partial_iA_j+\frac{\mu e}{mc}S_j\partial_iB_j,\label{nr9}
\\
\dot\omega_i=-\frac{2a^2}{b^2\phi}\pi_i+\frac{\mu e}{mc}\epsilon_{ijk}\omega_jB_k,
\qquad
\dot\pi_i=\frac{2}{\phi}\omega_i+\frac{\mu e}{mc}\epsilon_{ijk}\pi_jB_k.\nonumber
\end{gather}
The variable $S_i=\epsilon_{ijk}\omega_j\pi_k$ have unambiguous evolution, as it should be
\begin{gather*}
\dot S_i=\frac{\mu e}{mc}\epsilon_{ijk}S_jB_k.
\end{gather*}
This is the classical equation for precession of spin in an external magnetic f\/ield.
Due to equations~\eqref{intr.11}, the coordinates $S_i$ obey~\eqref{intr.12}.
Equations~\eqref{nr9} imply the second-order equation for~$x_i$
\begin{gather}
\label{nr13}
m\ddot x_i=\frac{e}{c}\epsilon_{ijk}\dot x_jB_k+\frac{e}{mc}S_k\partial_iB_k.
\end{gather}
Since ${\bf{S}}^2\sim\hbar^2$, the $S$-term disappears from equation~\eqref{nr13} at the classical limit
$\hbar\rightarrow 0$.
Then equation~\eqref{nr13} reproduces the classical motion of charged particle subject to the Lorentz force.
Note that in the absence of interaction, the particle does not experience a~self-acceleration.

To construct quantum mechanics of the spinning particle, we follow Dirac prescription for quantization of
a~system with constraints.
The constraints $\boldsymbol{\omega}^2-a^2=0$ and $(\boldsymbol{\omega}\boldsymbol{\pi})=0$ represent the
second-class system,
\begin{gather}
\label{qq1}
\big\{\boldsymbol{\omega}^2-a^2,(\boldsymbol{\omega}\boldsymbol{\pi})\big\}=2\boldsymbol{\omega}^2\ne0,
\end{gather}
while Poisson brackets of
\begin{gather}
\label{qq1.1}
\pi_\phi=0,
\qquad
\boldsymbol{\pi}^2-b^2+\frac{b^2}{a^2}\big(\boldsymbol{\omega}^2-a^2\big)=0,
\end{gather}
with all constraints vanish, so, they are the f\/irst-class constraints.

Remind that a~theory with second-class constraints, say $\Phi_a=0$, $\{\Phi_a, \Phi_b\}=\triangle_{ab}$,
$\det\triangle_{ab}\ne 0$, can not be consistently quantized on the base of Poisson bracket.
Indeed, since in classical theory $\Phi_a=0$, one expects that the corresponding operators vanish on
physical states, $\hat\Phi_a\Psi=0$.
Quantizing the theory by means of the Poisson bracket, we obtain
$(\hat\Phi_a\hat\Phi_b-\hat\Phi_b\hat\Phi_a)\Psi=\triangle_{ab}\Psi$.
The left-hand side of this expression vanishes, but the right-hand side does not.
The problem is resolved by postulating commutators that resemble the Dirac bracket
\begin{gather*}
\{A,B\}_{\rm D}=\{A,B\}-\{A,\Phi_a\}\triangle^{-1}_{ab}\{\Phi_b,B\},
\end{gather*}
instead of the Poisson one.
Owing to the property $\{\Phi_a, A\}_{\rm D}=0$, quantum analog of the Dirac bracket is consistent with the
condition $\hat\Phi_a\Psi=0$.
For our case the Dirac bracket reads
\begin{gather*}
\{A,B\}_{\rm D}=\{A,B\}+\big\{A,\boldsymbol{\omega}^2\big\}\frac{1}{2\boldsymbol{\omega}^2}\{\boldsymbol{\omega}
\boldsymbol{\pi},B\}-(A\leftrightarrow B).
\end{gather*}
For the basic variables this gives
\begin{gather*}
\{\omega_i,\omega_j\}=0,
\qquad
\{\omega_i,\pi_j\}=\delta_{ij}-\frac{\omega_i\omega_j}{\boldsymbol{\omega}^2},
\qquad
\{\pi_i,\pi_j\}=-\frac{\omega_i\pi_j-\omega_j\pi_i}{\boldsymbol{\omega}^2}.
\end{gather*}

Due to the property $\{\Phi_a, A\}_{\rm D}=0$, second-class constraints can now be used before computing the
bracket.
So, we can omit the third term in the Hamiltonian~\eqref{nr6}.
For the physical variab\-les~$x_i$, $p_i$, $S_i$, the Dirac bracket coincides with the Poisson one
\begin{gather}
\label{qq7}
\{x_i,p_j\}_{\rm D}=\delta_{ij},
\qquad
\{S_i,S_j\}_{\rm D}=\epsilon_{ijk}S_k.
\end{gather}

Now we are ready to complete canonical quantization of the model.
We quantize only the physical variables.
As the last two terms in~\eqref{nr6} does not contributes into equations of motion for the physical
variables, we omit them.
This gives the physical Hamiltonian\footnote{Equivalently, we could impose the gauge $\phi=1$, $\omega_3=0$
for the f\/irst-class constraints~\eqref{qq1.1} and construct the corresponding Dirac bracket.
This does not spoil neither the brackets~\eqref{qq7} nor equations of motion for physical variables.
Dealing with the Dirac bracket, all the constraints can be omitted from equation~\eqref{nr6}, this
gives~\eqref{qq6}.}
\begin{gather}
\label{qq6}
H=\frac{1}{2m}\left(p_i-\frac{e}{c}A_i\right)^2-\frac{\mu e}{mc}B_iS_i.
\end{gather}

The f\/irst equation from~\eqref{qq7} implies the standard quantization of the variables $x$ and $p$, we
take $\hat x_i=x_i$, $\hat p_i=-i\hbar\partial_i$.
According to the second equation from~\eqref{qq7}, we look for the wave-function space which is
representation of the group ${\rm SO}(3)$.
Finite-dimensional irreducible representations of the group are numbered by spin $s$, which is related with
values of Casimir operator as follows: ${\bf{S}}^2\sim s(s+1)$.
Then equation~\eqref{intr.12} f\/ixes the spin $s=\frac12$, and $S_i$ must be quantized by $\hat
S_i=\frac{\hbar}{2}\sigma_i$.
The operators act on space of two-component complex spinors~$\Psi$.
Quantum Hamiltonian is obtained from equation~\eqref{qq6} replacing classical variables by the operators.
This yields the Pauli equation
\begin{gather*}
i\hbar\frac{\partial\Psi}{\partial t}=\left(\frac{1}{2m}\left(\hat{\bf{p}}-\frac{e}{c}{\bf{A}}\right)^2-\frac{\mu e}
{mc}{\bf{B}}\hat{\bf{S}}\right)\Psi.
\end{gather*}
In resume, we have constructed non relativistic spinning particle with desired properties on both the
classical and the quantum level.

\section{Lorentz covariant form of spin f\/iber bundle}

In this section we represent ${\rm SO}(3)$ spin f\/iber bundle of Section~\ref{subsec:so3FB} in
the Lorentz-covariant form.
This yields automatically a~set of constraints which underly those determining Frenkel and BMT models.
We remind that our construction involves basic and target spaces: f\/iber bundle is a~submanifold of the
basic space, the base of the f\/iber bundle is a~submanifold of the target space.
First, we extend basic and target spaces in the following way.

Let $\Lambda: \omega'\rightarrow\Lambda\omega'$ be vector representation of the Lorentz group ${\rm SO}(1,
3)$, we introduce diagonal action of the group on space of direct product
$\mathbb{R}^8=\mathbb{R}^{1,3}\times \mathbb{R}^{1,3}$
\begin{gather}
\label{lf1}
{\rm SO}(1,3):
\
\left(
\begin{matrix}
\omega'
\\
\pi'
\end{matrix}
\right)\rightarrow\left(
\begin{matrix}
\omega
\\
\pi
\end{matrix}
\right)=\left(
\begin{matrix}
\Lambda & 0
\\
0 & \Lambda
\end{matrix}
\right)\left(
\begin{matrix}
\omega'
\\
\pi'
\end{matrix}
\right).
\end{gather}
Consider also six-dimensional space $\mathbb{R}^6$ with coordinates ${\bf k'}$, ${\bf j'}$.
Lorentz group naturally acts on the space, it is suf\/f\/icient to arrange the coordinates into $4\times 4$
antisymmetric matrix
\begin{gather*}
J^{\mu\nu}[{\bf k},{\bf j}]=\left(
\begin{matrix}
0& k'_1& k'_2& k'_3
\\
-k'_1& 0& j'_3& -j'_2
\\
-k'_2& -j'_3& 0& j'_1
\\
-k'_3& j'_2& -j'_1& 0
\end{matrix}
\right),
\end{gather*}
then the transformation
\begin{gather}
\label{def-act-SO31-R6}
{\rm SO}(1,3):
\
J^{\mu\nu}[{\bf k'},{\bf j'}]
\quad
\rightarrow
\quad
J^{\mu\nu}[{\bf k},{\bf j}]=\Lambda^{\mu}{}_\alpha\Lambda^{\nu}{}_\beta J^{\alpha\beta}[{\bf k'},{\bf j'}].
\end{gather}
determines transformation rules of ${\bf k'}$ and ${\bf j'}$.
Next we def\/ine a~map $f$ from $\mathbb{R}^{1,3}\times \mathbb{R}^{1,3}$ into $\mathbb{R}^6$
\begin{gather}
\label{def:projection-map-so13}
f:
\
\big(\omega'^\mu,\pi'^\nu\big)
\quad
\rightarrow
\quad
J^{\mu\nu}[{\bf k'},{\bf j'}]=2(\omega'^\mu\pi'^\nu-\omega'^\nu\pi'^\mu).
\end{gather}
This map has rank equals $5$, it maps a~point from $\mathbb{R}^{1,3}\times \mathbb{R}^{1,3}$ to a~pair of
orthogonal three-dimensional vectors, $({\bf k'},{\bf j'})=0$.

If $\mathbb{R}^8$ is considered as a~symplectic space with canonical Poisson bracket,
$\{\omega'^\mu,\pi'^\nu\}=\eta^{\mu\nu}$, the map $f$ induces ${\rm SO}(1, 3)$-Lie--Poisson bracket on~$\mathbb{R}^6$
\begin{gather*}
\big\{J^{\mu\nu}(\omega',\pi'),J^{\alpha\beta}(\omega',\pi')\big\}
=2\big(\eta^{\mu\alpha}J^{\nu\beta}-\eta^{\mu\beta}J^{\nu\alpha}-\eta^{\nu\alpha}J^{\mu\beta}+\eta^{\nu\beta}J^{\mu\alpha}\big).
\end{gather*}
By construction, $f$ is compatible with actions~\eqref{lf1} and~\eqref{def-act-SO31-R6} of ${\rm SO}(1,3)$:
if $J^{\mu\nu}[{\bf k'},{\bf j'}]$ $=2(\omega'^\mu \pi'^\nu-\omega'^\nu\pi'^\mu)$, then $J^{\mu\nu}[{\bf
k},{\bf j}] =2(\omega^\mu\pi^\nu-\omega^\nu\pi^\mu)$.

\subsection[Spin f\/iber bundle $\mathbb{T}^3$]{Spin f\/iber bundle $\boldsymbol{\mathbb{T}^3}$}

Given a~particular coordinate system $(\omega'^\mu,\pi'^\nu)$ in the basic space, let us consider the surface
\begin{gather}
\label{complete-spin-surface-rest-frame}
\boldsymbol{\pi}'^2=a_3,
\qquad
\boldsymbol{\omega}'^2=a_4,
\qquad
\boldsymbol{\omega}'\boldsymbol{\pi}'=0,
\qquad
\pi'_0=0,
\qquad
\omega'_0=0.
\end{gather}
Comparing this with~\eqref{ss.4} we identify the spin f\/iber bundle~\eqref{ss.7}, $\mathbb{T}^3\sim {\rm
SO}(3)$, with this surface.
Being restricted to the surface~\eqref{complete-spin-surface-rest-frame}, the map $f$ reads
\begin{gather}
\label{def:rest-frame-J}
J^{\mu\nu}|_{f(\mathbb{T}^3)}=\left(
\begin{matrix}
0& 0& 0& 0
\\
0& 0& j'_3& -j'_2
\\
0& -j'_3& 0& j'_1
\\
0& j'_2& -j'_1& 0
\end{matrix}
\right),
\qquad
{\bf j}'=2\boldsymbol\omega'\times\boldsymbol\pi'.
\end{gather}
Comparing equations~\eqref{def:projection-map-so13}
and~\eqref{complete-spin-surface-rest-frame}--\eqref{def:rest-frame-J} with~\eqref{ss.2} and~\eqref{ss.4}
we conclude that ${\rm SO}(3)$-construction of previous section is embedded into ${\rm SO}(1, 3)$.

To write the spin f\/iber bundle~\eqref{complete-spin-surface-rest-frame} in a~manifestly covariant form,
we see how this surface looks like in the coordinate system related with $(\omega', \pi')$ by ${\rm SO}(1,
3)$-transformation.
Since the surface is already invariant under ${\rm SO}(3)\subset {\rm SO}(1,3)$, we consider only boost to
the system moving with velocity ${\bf v}$.
Using the standard notation for relativistic factors $\boldsymbol{\beta}=\frac{{\bf v}}{c}$ and
$\gamma=(1-{\boldsymbol\beta}^2)^{-1/2}$ we have
\begin{gather}
\label{xx1}
\left(
\begin{matrix}
\omega'^0
\\
\boldsymbol{\omega}'
\end{matrix}
\right)=\left(
\begin{matrix}
\gamma & \gamma\boldsymbol{\beta}^{\rm T}
\vspace{1mm}\\
\gamma\boldsymbol{\beta} & I_3+\dfrac{\gamma-1}{\beta^2}\boldsymbol{\beta}\boldsymbol{\beta}^{\rm T}
\end{matrix}
\right)\left(
\begin{matrix}
\omega^0
\\
\boldsymbol{\omega}
\end{matrix}
\right),
\end{gather}
Applying the boost to~\eqref{complete-spin-surface-rest-frame} we obtain covariant equations of the surface
$\mathbb{T}^3$ in an arbitrary reference frame
\begin{gather}
T_3=\pi^2-a_3=0,
\qquad
T_4=\omega^2-a_4=0,
\qquad
T_5=\omega\pi=0,\nonumber
\\
T_6={\cal P}\omega=0,
\qquad
T_7={\cal P}\pi=0.
\label{def:complete-spin-surface}
\end{gather}
We have introduced time-like four-vector
\begin{gather*}
{\cal P}^\mu=(\tilde m\gamma c,\tilde m\gamma{\bf v}),
\qquad
\text{then}
\qquad
\gamma=\frac{|{\cal P}^0|}{\sqrt{{\cal P}_0^2-{\boldsymbol{\cal P}}^2}},
\qquad
\boldsymbol\beta=\frac{\boldsymbol{{\cal P}}}{{\cal P}^0}.
\end{gather*}
{}$\tilde m$ is an ef\/fective mass related with this vector
\begin{gather*}
\tilde m c=\sqrt{{\cal P}_0^2-\boldsymbol{{\cal P}}^2}.
\end{gather*}
In the dynamical model this vector appeared as the four-momentum of spinning particle.

\begin{figure}[t]\centering
\includegraphics[width=5.5in]{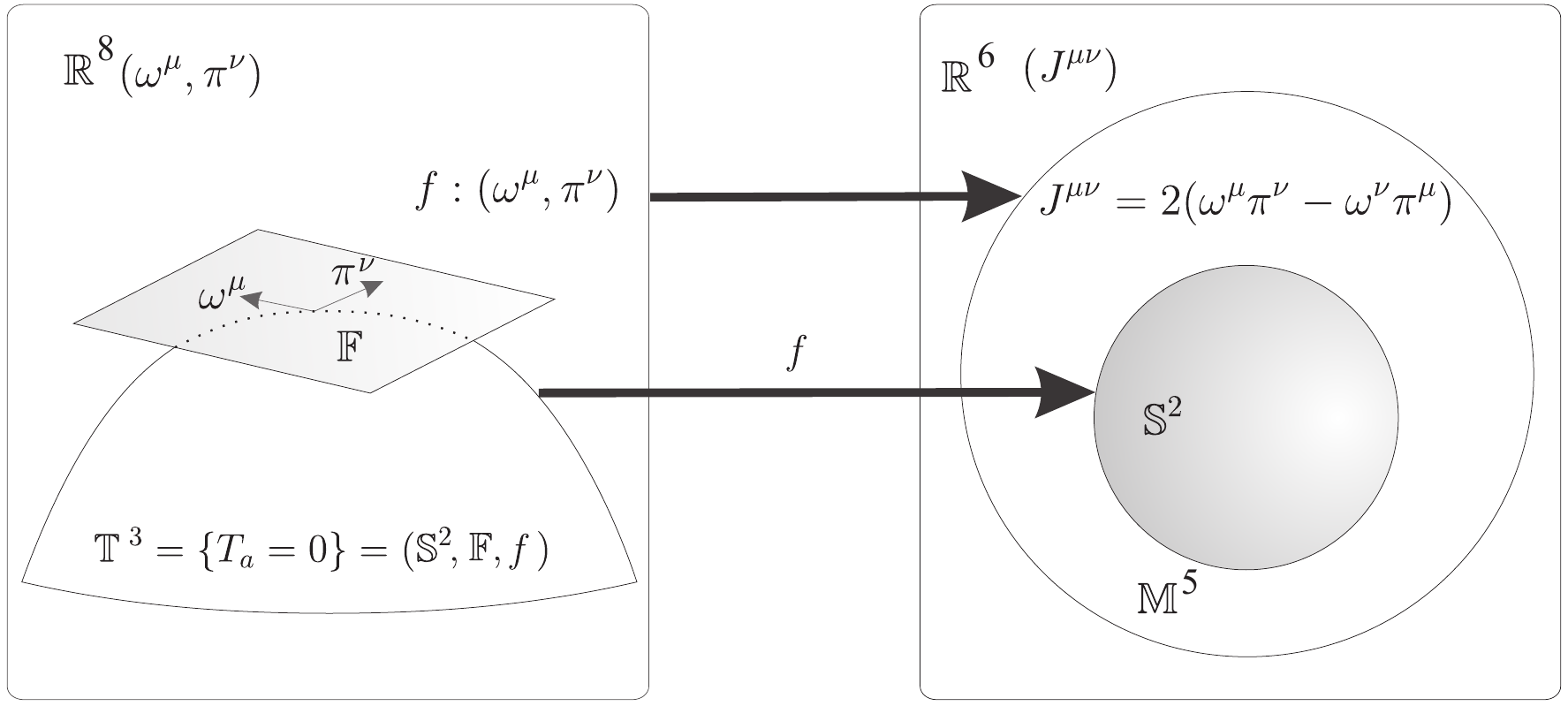}
\caption{Identif\/ication of spin surface
$\mathbb{S}^{2}$ with base of the spin f\/iber bundle $\mathbb{T}^{3}$.}
\label{fig:spin-fiber-bundle}
\end{figure}

Equations~\eqref{def:complete-spin-surface} determine the spin f\/iber
bundle $\mathbb{T}^3\sim {\rm SO}(3)$ in a~covariant form.
Let us describe its structure (see Fig.~\ref{fig:spin-fiber-bundle}).
The covariant projection map has been already def\/ined by~\eqref{def:projection-map-so13}, its form is
independent from the choice of coordinates.
The image of $f(\mathbb{T}^3)\subset \mathbb{R}^6$ is a~base of~$\mathbb{T}^3$.
Due to equations~\eqref{def:complete-spin-surface}, this is given by the
following 5 equations
\begin{gather}
\label{def:invariant-J1-on-T3}
J_{\mu\nu}J^{\mu\nu}=8\left[(T_4-a_4)(T_3-a_3)-T_5^2\right]
\qquad
\Rightarrow
\qquad
J_{\mu\nu}J^{\mu\nu}|_{\mathbb{T}^3}=8a_3a_4,
\\
\label{def:Shirokov-cond}
J^{\mu\nu}{\cal P}_\nu=0.
\end{gather}
The last equation represents the Frenkel-type condition necessary for construction of Frenkel and BMT
equations.

As $(J^{\mu\nu}{\cal P}_\mu){\cal P}_\nu\equiv 0$, we have 4 independent equations imposed on 6 variables,
therefore the base has dimension 2, as it should be.
Denote $\mathbb{F}_S\in\mathbb{T}^3$ preimage of a~point $J$ of the base, $\mathbb{F}_J=f^{-1}(J)$.
This set is composed by all pairs $(\omega, \pi)$ which lie on the same plane and thus related by ${\rm
SO}(2)$ rotations of the plane.
In the result, the manifold $\mathbb{T}^3$ acquires structure of f\/iber bundle with the base determined by
equations~\eqref{def:invariant-J1-on-T3} and~\eqref{def:Shirokov-cond}, standard f\/iber $\mathbb{F}$,
projection map $f$ and structure group ${\rm SO}(2)$.

Let us determine a~conventional coordinate system of the base.
We write independent among the equations~\eqref{def:invariant-J1-on-T3} and~\eqref{def:Shirokov-cond}
explicitly in the vector form
\begin{gather*}
{\bf k}=\frac{1}{{\cal P}^0}[{\bf j}\times\boldsymbol{{\cal P}}],
\qquad
\big({\bf j}^2-{\bf k}^2\big)=3\hbar^2.
\end{gather*}
From these equations follows that ${\bf k}$ is orthogonal to the plane def\/ined by ${\bf j}$ and
$\boldsymbol{{\cal P}}$.
Exclu\-ding~${\bf k}$ we obtain a~single equation of a~quadric surface
\begin{gather*}
{\bf j}^2-\frac{1}{{\cal P}_0^2}[{\bf j}\times\boldsymbol{{\cal P}}]^2=3\hbar^2.
\end{gather*}
Consider a~spherical coordinate system in $\mathbb{R}^{3}$ with the zenith direction given by vector
$\boldsymbol{{\cal P}}$.
The spherical coordinates of vectors ${\bf j}$ and ${\bf k}$ read
\begin{gather*}
{\bf j}=\left(j,\theta,\phi\right),
\qquad
{\bf k}=\left(k,\frac{\pi}{2},\phi+\frac{\pi}{2}\right),
\end{gather*}
where $\theta$ and $\phi$ are polar and azimuthal angles of ${\bf j}$.
In the spherical coordinates the equation of base reads
\begin{gather*}
j^2\big(1-\sin^2(\theta)\beta^2\big)=3\hbar^2.
\end{gather*}
Returning to Cartesian coordinates with the third axis along $\boldsymbol{{\cal P}}$, we recognize the
equation of an ellipsoid
\begin{gather*}
j_3^2+\frac{j_2^2+j_1^2}{\gamma^2}=3\hbar^2,
\end{gather*}
that is $f(\mathbb{T}^3)\sim S^2$.
The lengths of ${\bf j}$ and ${\bf k}$ are
\begin{gather*}
j=\hbar\sqrt{\frac{3}{1-\sin^2(\theta)\beta^2}},
\qquad
k=\hbar\beta\sin(\theta)\sqrt{\frac{3}{1-\sin^2(\theta)\beta^2}}.
\end{gather*}
In the rest coordinate system, when $\boldsymbol{\beta}=0$ this ellipsoid turns into a~sphere.

Finally we introduce BMT four-vector of spin which generalizes~\eqref{ss.2}
\begin{gather*}
S^\mu=\frac{1}{\tilde m c}\epsilon^{\mu\nu\alpha\beta}{\cal P}_\nu\omega_{\alpha}\pi_\beta=\frac{1}
{4\sqrt{-{\cal P}^2}}\epsilon^{\mu\nu\alpha\beta}{\cal P}_\nu J_{\alpha\beta}.
\end{gather*}
{}$S^\mu$ can be expressed in terms of ${\bf j}$ and vise a~versa
\begin{gather*}
S^0=\frac{\gamma}{2}(\boldsymbol{\beta}{\bf j}),
\qquad
{\bf S}=\frac{1}{2}\left(\frac{1}{\gamma}{\bf j}
+\gamma\boldsymbol{\beta}(\boldsymbol{\beta}{\bf j})\right),
\qquad
{\bf j}=2\gamma({\bf S}-\boldsymbol{\beta}(\boldsymbol{\beta}{\bf S})),
\qquad
{\bf k}=2\gamma[{\bf S}\times\boldsymbol{\beta}].
\end{gather*}

Four mutually orthogonal vectors $\omega^\mu$, $\pi^\mu$, ${\cal P}^\mu$ and $S^\mu$ allow us to def\/ine
uniquely a~basis in~$\mathbb{R}^{1,3}$.
They determine also the following element of ${\rm SO}(1,3)$
\begin{gather*}
\Lambda^{\mu}{}_{\nu}=\left(
\begin{matrix}
(\tilde m c)^{-1}{\cal P}^\mu
\\
a_4^{-1/2}\omega^\mu
\\
a_3^{-1/2}\pi^\mu
\\
(2a_3a_4)^{-1/2}S^\mu
\\
\end{matrix}
\right),
\qquad
\Lambda^{\mu\alpha}\eta_{\alpha\beta}\Lambda^{\nu\beta}=\eta^{\mu\nu}.
\end{gather*}
This is the element~\eqref{ss.5} of ${\rm SO}(3)$-group written in the boosted frame~\eqref{xx1}.

\subsection[Spin f\/iber bundle $\mathbb{T}^4$]{Spin f\/iber bundle $\boldsymbol{\mathbb{T}^4}$}

Consider the set of Lorentz-covariant constraints
\begin{gather}
\label{uf4.14}
{\cal P}\omega=0,
\qquad
{\cal P}\pi=0,
\\
\label{uf4.14.1}
\omega\pi=0,
\qquad
\pi^2-\frac{a}{\omega^2}=0.
\end{gather}
To see their meaning, we pass to the rest frame of ${\cal P}^\mu$, that is ${\cal P}^\mu=({\cal P}^0, {\bf 0})$.
Then equations~\eqref{uf4.14} mean $\omega^0=\pi^0=0$.
Taking this into account, the remaining constraints determines the following surface in $\mathbb
{R}^6(\boldsymbol\omega, \boldsymbol\pi)$
\begin{gather}
\label{uf4.15}
\mathbb{T}^4=\left\{\boldsymbol\omega\boldsymbol\pi=0,\,
\boldsymbol\pi^2-\frac{a}{\boldsymbol\omega^2}=0\right\},
\end{gather}
that is $\boldsymbol\omega$ and $\boldsymbol\pi$ represent a~pair of orthogonal vectors with ends attached
to hyperbole $y=\frac{a}{x}$.
Besides, the constraints~\eqref{uf4.14} imply $J^{\mu\nu}{\cal P}_\nu=0$.
In the rest frame this gives $J^{i0}=0$, that is the spin-tensor has only three components which we
identify with non-relativistic spin-vector, $J_{ij}=\epsilon_{ijk}S_k$.
The constraints~\eqref{uf4.15} then imply that the spin-vector belong to two-dimensional sphere of radius~$\sqrt{a}$
\begin{gather*}
J_{ij}J_{ij}=8a,
\qquad
\text{or}
\qquad
{\bf S}^2=a,
\qquad
\text{so we assume}
\quad
a=\frac{3\hbar^2}{4}.
\end{gather*}
The chosen value of parameter corresponds to spin one-half particle.

Hence, to describe spin in the rest frame, we have six-dimensional space of basic variables $\mathbb
{R}^6(\boldsymbol\omega, \boldsymbol\pi)$, the spin-tensor space $\mathbb {R}^3(J_{ij}\sim{\bf S})$ and the
map
\begin{gather*}
f: \ \mathbb{R}^{6}~\rightarrow~\mathbb{R}^{3},
\qquad
f: \ (\boldsymbol\omega,\boldsymbol\pi)~\rightarrow~{\bf S}=\boldsymbol\omega\times\boldsymbol\pi,
\qquad
\text{rank}\frac{\partial(S_i)}{\partial(\omega_j,\pi_k)}=3.
\end{gather*}
 $f$ maps the manifold $\mathbb{T}^4$ onto spin surface, $f(\mathbb{T}^4)=\mathbb{S}^2$.

Denote $\mathbb{F}^2_S\in\mathbb{T}^4$ preimage of a~point ${\bf{S}}\in\mathbb{S}^2$,
$\mathbb{F}^2_S=f^{-1}({\bf{S}})$.
Let $(\boldsymbol{\omega}, \boldsymbol{\pi})\in\mathbb{F}^2_S$.
Then the two-dimensional manifold $\mathbb{F}^2_S$ contains all pairs $(k\boldsymbol\omega,
\frac{1}{k}\boldsymbol\pi)$, $k\in\mathbb{R^{+}}$, as well as the pairs obtained by rotation of these
$(k\boldsymbol\omega, \frac{1}{k}\boldsymbol\pi)$ in the plane of vectors $(\boldsymbol{\omega},\boldsymbol{\pi})$.
So elements of $\mathbb{F}^2_S$ are related by two-parametric transformations
\begin{gather}
\label{uf4.18}
\boldsymbol{\omega}'=\boldsymbol{\omega}k\cos\beta+\boldsymbol{\pi}\frac{k|\boldsymbol\omega|}
{|\boldsymbol\pi|}\sin\beta,
\qquad
\boldsymbol{\pi}'=-\boldsymbol{\omega}\frac{|\boldsymbol\pi|}{k|\boldsymbol\omega|}
\sin\beta+\boldsymbol{\pi}\frac{1}{k}\cos\beta.
\end{gather}
In the result, the manifold $\mathbb{T}^4$ acquires natural structure of f\/iber bundle
\begin{gather*}
\mathbb{T}^4=\big(\mathbb{S}^2,\mathbb{F}^2,f\big),
\end{gather*}
with base $\mathbb{S}^2$, standard f\/iber $\mathbb{F}^2$, projection map $f$ and structure group given by
transformations~\eqref{uf4.18}.
As local coordinates of $\mathbb{T}^4$ adjusted with the structure of f\/iber bundle we can take~$k$,~$\beta$, and two coordinates of the vector~${\bf S}$.
By construction, the structure-group transformations leave inert points of base, $\delta S_i=0$.

The covariant equations~\eqref{uf4.14}--\eqref{uf4.14.1} together with the map
$J^{\mu\nu}=2\omega^{[\mu}\pi^{\nu]}$ represent this construction in an arbitrary Lorentz frame.

\section{Conclusion}

Reparametrization symmetry is known to be crucial for Lorentz-covariant description of a~spinless particle.
To describe a~spinning particle on the base of vector-like variable, we need one more local symmetry
written in equation~\eqref{nnr3.2}.
The spin-plane symmetry appears already in non-relativistic model.
Together with two second-class constraints~\eqref{qq1}, this guarantees the right number of degrees of
freedom and determines physical sector of the model.
Variational formulation of the spinning particle implies a~singular Lagrangian which leads to a~curved
phase-space endowed with the structure of f\/iber bundle~\eqref{ss.7}.
The local symmetry~\eqref{nnr3.2} represents transformations of structure group~\eqref{ss.8} acting
independently at each instance of time, and has clear geometric interpretation: this corresponds to
rotations of the pair ${\boldsymbol\omega}$, ${\boldsymbol\pi}$ in the plane formed by these vectors.
Equation~\eqref{ss.10} suggests that the matrix~\eqref{ss.11}, formed from auxiliary variables of the
model, play a~role of gauge f\/ield associated with the symmetry.

We have described the spin f\/iber bundles~\eqref{intr.11} and~\eqref{intr12.1} in Lorentz-covariant form,
$\mathbb{T}^3$~and~$\mathbb{T}^4$.
The resulting sets of covariant
constraints~\eqref{def:complete-spin-surface}
and~\eqref{uf4.14}--\eqref{uf4.14.1} guarantee the Frenkel-type condition~\eqref{def:Shirokov-cond}, so
dynamical realization of $\mathbb{T}^3$ gives variational formulation for the Frenkel and BMT
equations~\cite{Alexei, DPW2}.
The constraints~\eqref{uf4.14.1} of the space $\mathbb{T}^4$ appeared in the model of rigid
particle~\cite{deriglazov2013rigid}.

In non-relativistic model~\eqref{nr3}, the spin-plane symmetry acts only in the spin-sector.
Rela\-ti\-vis\-tic description of spin can lead to nontrivial transformation law of the position
variable~$x^\mu$~\cite{AAD5}.
This turns out to be crucial point for various issues including the identif\/ication of operators of the
Dirac equation with their classical analogs, and analysis of {\it Zitterbewegung} phenomenon.
For instance, the model discussed in~\cite{AAD6} indicates that the {\it Zitterbewegung} represents an
evolution of gauge non-invariant (hence unobservable) variables.

\subsection*{Acknowledgments} The work of AAD has been supported by the Brazilian foundation CNPq.
AMPM thanks CAPES for the f\/inancial support (Programm PNPD/2011).

\pdfbookmark[1]{References}{ref}
\LastPageEnding


\begin{thebibliography}{99}
\footnotesize\itemsep=0pt

\bibitem{BMT}
Bargmann V., Michel L., Telegdi V.L., Precession of the polarization of
  particles moving in a homogeneous electromagnetic f\/ield, \href{http://dx.doi.org/10.1103/PhysRevLett.2.435}{\textit{Phys. Rev.
  Lett.}} \textbf{2} (1959), 435--436.

\bibitem{bbra}
Barut A.O., Bracken A.J., Zitterbewegung and the internal geometry of the
  electron, \href{http://dx.doi.org/10.1103/PhysRevD.23.2454}{\textit{Phys. Rev.~D}} \textbf{23} (1981), 2454--2463.

\bibitem{bt}
Barut A.O., Thacker W., Covariant generalization of the {Z}itterbewegung of the
  electron and its {${\rm SO}(4,2)$} and {${\rm SO}(3,2)$} internal algebras,
  \href{http://dx.doi.org/10.1103/PhysRevD.31.1386}{\textit{Phys. Rev.~D}} \textbf{31} (1985), 1386--1392.

\bibitem{berezin:1977}
Berezin F.A., Marinov M.S., Particle spin dynamics as the Grassmann variant of
  classical mechanics, \href{http://dx.doi.org/10.1016/0003-4916(77)90335-9}{\textit{Ann. Physics}} \textbf{104} (1977), 336--362.

\bibitem{cognola1981lagrangian}
Cognola G., Vanzo L., Zerbini S., Soldati R., On the Lagrangian formulation of
  a charged spinning particle in an external electromagnetic f\/ield,
  \href{http://dx.doi.org/10.1016/0370-2693(81)90856-X}{\textit{Phys. Lett.~B}} \textbf{104} (1981), 67--69.

\bibitem{corben:1968}
Corben H.C., Classical and quantum theories of spinning particles, Holden-Day,
  San Francisco, 1968.

\bibitem{deriglazov2010classical}
Deriglazov A.A., Classical mechanics: Hamiltonian and Lagrangian formalism,
  \href{http://dx.doi.org/10.1007/978-3-642-14037-2}{Springer}, Heidelberg, 2010.

\bibitem{deriglazov-ns:2010}
Deriglazov A.A., Nonrelativistic spin: \`a la {B}erezin--{M}arinov quantization
  on a sphere, \href{http://dx.doi.org/10.1142/S0217732310033980}{\textit{Modern Phys. Lett.~A}} \textbf{25} (2010), 2769--2777.

\bibitem{AAD3}
Deriglazov A.A., A semiclassical description of relativistic spin without the
  use of {G}rassmann variables and the {D}irac equation, \href{http://dx.doi.org/10.1016/j.aop.2011.11.019}{\textit{Ann. Physics}}
  \textbf{327} (2012), 398--406, \href{http://arxiv.org/abs/1107.0273}{arXiv:1107.0273}.

\bibitem{AAD6}
Deriglazov A.A., Classical-mechanical models without observable trajectories
  and the {D}irac electron, \href{http://dx.doi.org/10.1016/j.physleta.2012.11.024}{\textit{Phys. Lett.~A}} \textbf{377} (2012), 13--17,
  \href{http://arxiv.org/abs/1203.5697}{arXiv:1203.5697}.


\bibitem{AAD5}
Deriglazov A.A., Spinning-particle model for the {D}irac equation and the
  relativistic {Z}itterbewegung, \href{http://dx.doi.org/10.1016/j.physleta.2011.10.070}{\textit{Phys. Lett.~A}} \textbf{376} (2012),
  309--313, \href{http://arxiv.org/abs/1106.5228}{arXiv:1106.5228}.

\bibitem{Alexei}
Deriglazov A.A., Variational problem for the {F}renkel and the
  {B}argmann--{M}ichel--{T}elegdi ({BMT}) equations, \href{http://dx.doi.org/10.1142/S0217732312502343}{\textit{Modern Phys.
  Lett.~A}} \textbf{28} (2013), 1250234, 9~pages, \href{http://arxiv.org/abs/1204.2494}{arXiv:1204.2494}.

 \bibitem{deriglazov2012variational}
Deriglazov A.A., Variational problem for Hamiltonian system on ${\rm SO}(k,m)$
  {L}ie--{P}oisson manifold and dynamics of semiclassical spin,
  \href{http://arxiv.org/abs/1211.1219}{arXiv:1211.1219}.

\bibitem{deriglazov2000local}
Deriglazov A.A., Evdokimov K.E., Local symmetries and the {N}oether identities
  in the {H}amiltonian framework, \href{http://dx.doi.org/10.1142/S0217751X00001899}{\textit{Internat.~J. Modern Phys.~A}}
  \textbf{15} (2000), 4045--4067, \href{http://arxiv.org/abs/hep-th/9912179}{hep-th/9912179}.

\bibitem{deriglazov2013rigid}
Deriglazov A.A., Nersessian A., Rigid particle revisited: extrinsic curvature
  yields the {D}irac equation, \href{http://arxiv.org/abs/1303.0483}{arXiv:1303.0483}.

\bibitem{DPW1}
Deriglazov A.A., Pupasov-Maksimov A.M., Lagrangian for Frenkel electron and
  position's non-com\-mu\-ta\-ti\-vi\-ty due to spin, \href{http://arxiv.org/abs/1312.6247}{arXiv:1312.6247}.

\bibitem{AAD4}
Deriglazov A.A., Rizzuti B.F., Zamudio G.P., Castro P.S., Non-{G}rassmann
  mechanical model of the {D}irac equation, \href{http://dx.doi.org/10.1063/1.4759500}{\textit{J.~Math. Phys.}} \textbf{53}
  (2012), 122303, 31~pages, \href{http://arxiv.org/abs/1202.5757}{arXiv:1202.5757}.

\bibitem{dirac1950lectures}
Dirac P.A.M., Lectures on quantum mechanics, \textit{Belfer Graduate School of
  Science Monographs Series}, Vol.~2, Belfer Graduate School of Science, New
  York, 1967.

\bibitem{foldy:1978}
Foldy L.L., Wouthuysen S.A., On the {D}irac theory of spin~1/2 particles and
  its non-relativistic limit, \href{http://dx.doi.org/10.1103/PhysRev.78.29}{\textit{Phys. Rev.}} \textbf{78} (1950), 29--36.

\bibitem{Frenkel}
Frenkel J., Die Elektrodynamik des rotierenden Elektrons, \href{http://dx.doi.org/10.1007/BF01397099}{\textit{Z.~Phys.}}
  \textbf{37} (1926), 243--262.

\bibitem{Frenkel2}
Frenkel J., Spinning electrons, \href{http://dx.doi.org/10.1038/117653a0}{\textit{Nature}} \textbf{117} (1926), 653--654.

\bibitem{GG1}
Gavrilov S.P., Gitman D.M., Quantization of pointlike particles and consistent
  relativistic quantum mechanics, \href{http://dx.doi.org/10.1142/S0217751X00002376}{\textit{Internat.~J. Modern Phys.~A}}
  \textbf{15} (2000), 4499--4538, \href{http://arxiv.org/abs/hep-th/000311}{hep-th/0003112}.

\bibitem{gitman1990quantization}
Gitman D.M., Tyutin I.V., Quantization of f\/ields with constraints, \href{http://dx.doi.org/10.1007/978-3-642-83938-2}{\textit{Springer
  Series in Nuclear and Particle Physics}}, Springer-Verlag, Berlin, 1990.

\bibitem{grassberger1978}
Grassberger P., Classical charged particles with spin, \href{http://dx.doi.org/10.1088/0305-4470/11/7/009}{\textit{J.~Phys.~A:
  Math. Gen.}} \textbf{11} (1978), 1221--1226.

\bibitem{hanson1974relativistic}
Hanson A.J., Regge T., The relativistic spherical top, \href{http://dx.doi.org/10.1016/0003-4916(74)90046-3}{\textit{Ann. Physics}}
  \textbf{87} (1974), 498--566.

\bibitem{laroze2009spin-gravity}
Laroze D., Guti{\'e}rrez G., Rivera R., Y{\'a}{\~n}ez J.M., Dynamics of a
  rotating particle under a time-dependent potential: exact quantum solution
  from the classical action, \href{http://dx.doi.org/10.1088/0031-8949/78/01/015009}{\textit{Phys. Scr.}} \textbf{78} (2008), 015009,
  5~pages.

\bibitem{peletminski2005lagrangianspin}
Peletminskii A., Peletminskii S., Lagrangian and Hamiltonian formalisms for
  relativistic dynamics of a~charged particle with dipole moment, \href{http://dx.doi.org/10.1140/epjc/s2005-02336-4}{\textit{Eur.
  Phys.~J.~C Part. Fields}} \textbf{42} (2005), 505--517.

\bibitem{DPW2}
Ramirez W.G., Deriglazov A.A., Pupasov-Maksimov A.M., Frenkel electron and a
  spinning body in a curved background, \href{http://arxiv.org/abs/1311.5743}{arXiv:1311.5743}.

\end{thebibliography}
\end{document}